\documentclass[10pt,journal,a4paper]{IEEEtran}

\usepackage{color}
\usepackage{cite}      
\usepackage[caption=false, font=footnotesize]{subfig}
\usepackage{graphicx}
\usepackage[latin1]{inputenc}
\usepackage{amsfonts}
\usepackage{amsmath}
\usepackage{times}
\usepackage{algpseudocode}
\usepackage{algorithm}
\usepackage{enumerate}
\usepackage{siunitx}
\usepackage{multirow}
\usepackage{tikz}

\usepackage{amssymb}

\newtheorem{lemma}{Lemma}[section]
\newtheorem{theorem}{Theorem}[section]

\newtheorem{remark}{Remark}[section]
\newtheorem{example}{Example}[section]

\begin{document}

\title{Reliability of Multicast under Random Linear Network Coding}

\author{Evgeny Tsimbalo\thanks{
This work is partially supported by the \mbox{VENTURER} Project, which is supported by Innovate UK under Grant Number 102202.

E. Tsimbalo is with The Telecommunications Research Laboratory of Toshiba Research Europe Ltd., Bristol, BS1 4ND, UK (email: evgeny.tsimbalo@toshiba-trel.com).

A. Tassi and R. J. Piechocki are with the Department of Electrical and Electronic Engineering, University of Bristol, Bristol, BS8 1UB, UK (email: \{A.Tassi, R.J.Piechocki\}@bristol.ac.uk).},
Andrea Tassi and Robert J. Piechocki}
\maketitle

\begin{abstract}
We consider a lossy multicast network in which the reliability is provided by means of Random Linear Network Coding. Our goal is to characterise the performance of such network in terms of the probability that a source message is delivered to all destination nodes. Previous studies considered coding over large finite fields, small numbers of destination nodes or specific, often impractical, channel conditions. In contrast, we focus on a general problem, considering arbitrary field size and number of destination nodes, as well as a realistic channel. We propose a lower bound on the probability of successful delivery, which is more accurate than the approximation commonly used in the literature. {In addition, we present a novel performance analysis of the systematic version of RLNC. The accuracy of the proposed performance framework is verified via extensive Monte Carlo simulations, where the impact of the network and code parameters are investigated. Specifically, we show that the mean square error of the bound for a ten-user network can be as low as $9 \cdot 10^{-5}$ for non-systematic RLNC.}
\end{abstract}


\section{Introduction\label{sec:Introduction}}

Reliability is a key performance metric in modern wireless multicast networks, in which a single transmitter, or a \emph{source} node, broadcasts to multiple receivers, or \emph{destination} nodes, also referred to as \emph{users}. Traditionally, the reliability in multicast networks is provided by Application Level Forward Error Correction (AL-FEC) \cite{Gomez-Barquero2009}, where coding is performed over packets rather than bits. AL-FEC is typically based on a digital fountain approach \cite{Byers2002} implemented, for instance, in Raptor codes \cite{Wang2016b}. These codes, however, operate efficiently only when the number of packets per block is large, which makes them prohibitive in applications where the block size is limited - for instance, due to delay constraints \cite{Magli2013}.

As an alternative to traditional fountain codes, the idea of combining packets using \emph{random} linear coefficients \cite{Ho2003}, also known as \emph{Random Linear Network Coding} (RLNC) \cite{Ho2006}, has attracted significant research interest. RLNC is based on the original concept of network coding proposed by R. Ahlswede~\emph{et al.}~\cite{Ahlswede2000} and is proved to be capacity-achieving for lossy multicast networks \cite{Lun2004, Lun2005}. In contrast with traditional fountain codes, schemes based on RLNC do not require a large block size \cite{Magli2013}.

In a multicast network operated under RLNC, the source node encodes an information message of $K$ packets by combining the packets using random coefficients belonging to a finite field \cite{Chatzigeorgiou2017}. An encoded packet is therefore associated with a vector of coding coefficients. Each user needs to collect $K$ linearly independent vectors of coding coefficients to be able to decode the source message, typically by means of Gaussian elimination. A key performance indicator of such network can be the probability that all users collect $K$ linearly independent vectors of coding coefficients, which will be referred to as \emph{probability of successful delivery}. As an alternative to traditional, non-systematic encoding, a systematic version of RLNC, in which the information message is transmitted first, was also proposed \cite{Shrader2009,Heide2009,Lucani2010}. As shown in \cite{Barros2009}, systematic RLNC can reduce decoding delay and complexity.

The traditional approach to the performance analysis of RLNC and multicast networks is to assume an infinite, or sufficiently large, field size \cite{Ho2006,Eryilmaz2007,Lucani2009a}, so that any $K$ vectors of coding coefficients are linearly independent with a high probability. {Under this assumption, the multicast network can be viewed as a set of independent unicast connections. While the assumption significantly simplifies the analysis, the field size is limited in practice~\cite{Pedersen2013, Ferreira2014}. As a consequence of that, the probability that the vectors of coding coefficients collected by a user are linearly \emph{dependent} can be non-negligible, even if their number is larger than $K$.} In addition, some vectors of coding coefficients can be received simultaneously by multiple users, giving rise to statistical correlation. As a result, with a finite field size, the multicast network cannot be approximated as a set of independent unicast connections, and the probability of successful delivery needs to be calculated jointly.

The analysis of multicast networks and codes with coefficients generated from a small field size has been also studied in the literature. The initial studies \cite{Lucani2009, Nistor2011} were based on Markov chain models, but due to complexity the number of users was limited to two. Another study \cite{Heide2009} applied a Markov chain model to a network with an arbitrary number of users, assuming that the users receive disjoint sets of packets. In \cite{Trullols-Cruces2011}, an \emph{exact} probability of successful delivery, valid for any field size, was derived for the simple case of a unicast, point-to-point connection. The result was extended to the systematic version of RLNC in \cite{Jones2015}. Following that, in \cite{Khan2015,Amjad2015}, a two-user multicast network was considered in the context of security and relay communication, respectively. Assuming a sufficiently high packet erasure rate (PER), so that the users are likely to receive disjoint sets of packets, the probability of successful delivery was approximated as a product of those corresponding to each individual user.
In contrast, an \emph{exact} expression for the probability of successful delivery for the same network, valid for any field size \emph{and} PER, was obtained in \cite{Tsimbalo2016a} for non-systematic RLNC, based on the rank analysis of structured random matrices \cite{Ferreira2013}.

To summarise, the previous studies on the performance of RLNC and multicast networks considered large finite fields, a limited number of users or specific (often impractical) channel conditions. In the cases when the exact formulation formulation was obtained, the analysis was limited to two users. Moreover, the existing studies focus mainly on traditional, non-systematic RLNC. To the best of our knowledge, there is no study in the literature that considers a general case of arbitrary field size, number of users and channel conditions for both non-systematic and systematic RLNC.

In this work, we address the limitations of the previous studies and provide the following contributions:
\begin{itemize}
\item In contrast with \cite{Ho2006,Eryilmaz2007,Lucani2009a,Khan2015,Amjad2015}, we calculate the probability that all users recovered the source message \emph{jointly}, taking into account a finite field size and commonly received packets. 
\item We generalise the analysis limited to a two-user network \cite{Nistor2011, Tsimbalo2016a} to an \emph{arbitrary} number of users, and derive a tight lower bound for the probability of successful delivery in the case of traditional, non-systematic RLNC. In contrast with \cite{Khan2015,Amjad2015}, the bound takes into account the correlation effect due to commonly received packets.
\item We also present a novel analysis for the systematic version of RLNC. We argue that the correlation effect is less profound than in the non-systematic case, and each user can be considered independently, even if the field size is small. We formulate the result explicitly and prove that it is a tight lower bound.  
\item We perform thorough benchmarking of the proposed bounds via extensive Monte Carlo simulation, where the effects of the number of users, the PER, the source message size and the field size are investigated. We demonstrate that the considered bounds are especially accurate under realistic channel conditions. In the non-systematic case, the derived bound provides a much closer approximation than the traditional bound used in the literature. {In particular, this holds true in those scenarios where users are spread across the coverage area of the transmitter and experience heterogeneous PERs.}
\item We provide an extensive study into the performance of multicast networks under RLNC and offer an insight into the selection of code parameters for various network configurations.  
\end{itemize} 

The remainder of the paper is organised as follows. Section~\ref{sec:System-model} describes the system model and provides the necessary background on multicast networks. The proposed theoretical framework is presented in Section~\ref{sec:Theory}, where the bounds for the probability of successful delivery are derived for both non-systematic and systematic versions of RLNC. In Section~\ref{sec:NumResults}, the proposed bounds are compared with simulated results and existing bounds. Section~\ref{sec:Conclusions} draws conclusions and highlights future research avenues.

\section{System Model and Background\label{sec:System-model}}

Consider a multicast network, in which a source node transmits to $L$ destination nodes, or users. Each of the $L$ links is assumed to be lossy and characterised by a PER $\epsilon_j$ for $j=1,\ldots,L$. Here, we assume that packet erasures occur as statistically independent events. The goal is to deliver a message of $K$ source packets to each user. It will be assumed that the $i$-th source packet $\mathbf{s}_i$, $i=1,\ldots,K$, is a column vector of elements from a finite field $\mathbb{F}_q$ of size $q$. The number of elements in vector $\mathbf{s}_i$ is equal to $\lceil t / \log_2 q \rceil$ \cite{ho2008book}, where $t$ denotes the packet length in bits, which is assumed to be the same for all packets.

Given $K$ source packets, the encoder generates $N\geq K$ \emph{coded packets} $\{\mathbf{c}_k\}_{k=1}^{N}$, each being a vector consisting of $\lceil t / \log_2 q \rceil$ elements from $\mathbb{F}_q$. Using the matrix notation, the encoding operation can be expressed as follows:
\begin{equation}
	[\mathbf{c}_1,\ldots,\mathbf{c}_N] = [\mathbf{s}_1,\ldots,\mathbf{s}_K]\cdot \mathbf{G},
\end{equation}
where $\mathbf{G}\in \mathbb{F}_q^{K\times N}$ is a $K\times N$ matrix of \emph{coding coefficients} generated uniformly at random from $\mathbb{F}_q$. In the case of systematic RLNC, the first $K$ columns of $\mathbf{G}$ form an $K\times K$ identity matrix. In this way, the first $K$ transmissions are the source packets, also referred to as \emph{systematic} packets, followed by $N-K$ coded, non-systematic packets. It is beyond the scope of the paper to address sparse implementations of RLNC~\cite{8248799}.

Due to packet erasures, each user will receive a subset of transmitted packets. Let $U_{j}\subseteq\{1,\ldots,N\}$ denote a set of indices of transmitted packets received by the $j$-th user, $j=1,\ldots,L$. Let also $m_{j}=\left|U_{j}\right|$ denote the number of packets received by the $j$-th user. It is assumed that all users have a knowledge of the coding coefficients associated with each received packet. This can be achieved by transmitting the coefficients or the seed used to generate them in the packet header \cite{Ho2006}. The $j$-th user can therefore construct an $m_j\times K$ matrix of coding coefficients, which is obtained from $\mathbf{G}$ by deleting the rows corresponding to lost packets. This matrix will be denoted as $\mathbf{C}_j$ and will be referred to as the \emph{coding matrix} of the $j$-th user. A user can recover the source message if its coding matrix is full rank. We now define the \emph{probability of successful delivery} $P_L(\boldsymbol{\epsilon})$ of an $L$-user multicast network with PERs $\boldsymbol{\epsilon}=(\epsilon_1,\ldots,\epsilon_L)$ as the probability that all users have successfully recovered the source message.

\begin{table}
\begin{centering}
\caption{Notation used throughout the paper.}
\ifCLASSOPTIONtwocolumn
\begin{tabular}{|c|m{6.5cm}|}
\else
\begin{tabular}{|c|m{10cm}|}
\fi
\hline 
Notation & Description\tabularnewline
\hline 
\hline
$L$ & Number of user forming a multicast network\\ 
\hline
$K$ & Number of packets forming an information message\\ 
\hline
$N$ & Number of coded packet transmissions\\ 
\hline
$q$ & Size of the finite field under consideration\\ 
\hline
$U_j$ & Set of indices of transmitted packets received by the $j$-th user\\ 
\hline
$m_j$ & Total number of packets received by the $j$-th user\\ 
\hline
$\mu$ & Random variable denoting the number of packets received simultaneously by all the users\\ 
\hline
$\theta_J$ & Random variable denoting the number of packets received simultaneously by a subset $J$ of $L$ users, \mbox{$1<|J|<L$}\\ 
\hline
$\boldsymbol\theta$ & Tuple of variables $\theta_J$ for all possible subsets $J$, \mbox{$1<|J|<L$}\\ 
\hline
$\mathbb{P}(m,K)$ & Probability that an $m \times K$ matrix of elements generated uniformly at random from $\mathbb{F}_q$ is full rank\tabularnewline
\hline 
$\mathbb{P}^{(i)}(m,K)$ & Probability that an $m \times K$ matrix of elements generated uniformly at random from $\mathbb{F}_q$ has rank $i$\tabularnewline
\hline 
$P(\epsilon)$ & Probability of successful delivery over a point-to-point link with PER $\epsilon$ for non-systematic RLNC\tabularnewline
\hline 
$P^*(\epsilon)$ & Probability of successful delivery over a point-to-point link with PER $\epsilon$ for systematic RLNC\tabularnewline
\hline 
$P_L(\boldsymbol{\epsilon})$ & Probability of successful delivery over a multicast network with $L \geq 2$ users and PERs $\boldsymbol{\epsilon}=(\epsilon_1,\ldots,\epsilon_L)$ for non-systematic RLNC\tabularnewline
\hline
$P_L^*(\boldsymbol{\epsilon})$ & Probability of successful delivery over a multicast network with $L \geq 2$ users and PERs $\boldsymbol{\epsilon}=(\epsilon_1,\ldots,\epsilon_L)$ for systematic RLNC\tabularnewline
\hline
\end{tabular}
\par\end{centering}
\label{tab:Notation}
\end{table}
 
The simplest case of a multicast network is a point-to-point link with a single user characterised by a PER $\epsilon$. For a non-systematic code characterised by $(N,K,q)$, the probability of successful delivery for such link is given by \cite{Amjad2015}:
\begin{equation}
	P(\epsilon) = \sum_{m=K}^{N} \binom{N}{m} (1-\epsilon)^{m} \epsilon^{N-m} \mathbb{P}(m,K).\label{eq:P_ptp}
\end{equation}
Here, $\mathbb{P}(m,K)$ is the probability that an $m \times K$ matrix of elements generated uniformly at random from $\mathbb{F}_q$ is full rank, which is given by \cite{Trullols-Cruces2011}:
\begin{equation}
	\mathbb{P}(m,K) = \prod_{i=0}^{K-1}(1-q^{i-m}).\label{eq:P_single_mat}
\end{equation}
It can be observed that \eqref{eq:P_ptp} can be thought of as a marginalisation of the rank of the user's coding matrix over the distribution of the number of rows $m$ in this matrix.

For a systematic $(N,K,q)$ code and a point-to-point link, the probability of successful delivery can be expressed as follows \cite{Jones2015}:
\setlength{\arraycolsep}{0.1em}
\begin{eqnarray}
	P^*(\epsilon)	&=& \sum_{m = K}^{N} (1 - \epsilon)^{m} \epsilon^{N - m} \sum_{h = h_{min}}^K \binom{K}{h} \binom{N-K}{m-h}\nonumber\\ 
				 	& & \cdot \mathbb{P}(m-h, K-h),\label{eq:P_ptp_sys}
\end{eqnarray}
\setlength{\arraycolsep}{5pt}%
where $h$ denotes a possible number of received systematic packets and $h_{min}$ is defined as $\max(0,m-N+K)$. Compared with \eqref{eq:P_ptp} for the non-systematic case, we observe that an additional marginalisation over the distribution of $h$ is required for the systematic code. In addition, the number of ways to select $m$ received packets out of $N$ is replaced with the number of ways to select $h$ systematic packets out of $K$ and $m-h$ non-systematic packets out of $N-K$. Given that the user receives $h$ systematic packets out of $m$, $h$ columns of its coding matrix will be linearly independent. Therefore, for the matrix to be full rank, the remaining $K-h$ columns formed by the non-systematic coding vectors should be linearly independent. The minimum value of $h$, $h_{min}$, is chosen as the difference between the total number of received packets $m$ and a maximum possible number of non-systematic packets, $\min(m,(N-K))$.

Consider now the general case of an $L$-user multicast network. As mentioned in Section~\ref{sec:Introduction}, if the field size $q$ is sufficiently large, each user is able to recover the message with a high probability once it receives at least $K$ packets. Indeed, \eqref{eq:P_single_mat} is close to $1$ for large $q$. In this case, the users will recover the message independently from each other and the probability of successful delivery can be approximated as follows (in the case of a non-systematic $(N,K,q)$ code):
\begin{equation}
	P_L(\boldsymbol{\epsilon}) \cong \prod_{j=1}^{L}P(\epsilon_j),\label{eq:P_M_approx}
\end{equation}
where $P(\epsilon_j)$ is the probability of successful delivery of a source message over a point-to-point link with a PER $\epsilon_j$ corresponding to the $j$-th user, as calculated by \eqref{eq:P_ptp}. It should be noted, however, that with a limited field size $q$, the accuracy of \eqref{eq:P_M_approx} is expected to decrease as the number of users grows.

For a specific case of $L=2$, it was shown in \cite{Amjad2015} that \eqref{eq:P_M_approx} is a good approximation even if $q$ is small, provided that the number of transmissions and PER are high enough for the users to receive independent subsets of packets. By contrast, an exact formulation for the probability of successful delivery, valid for any field size, number of transmissions and channel conditions, was obtained for a two-user multicast network  in \cite{Tsimbalo2016a} for non-systematic RLNC. The exact formulation is given as follows:
\setlength{\arraycolsep}{0.14em} 
\begin{eqnarray}
	\hspace{-1em}P_2(\boldsymbol{\epsilon}) & = & \sum_{m_1=K}^{N} \sum_{m_2=K}^{N} \left(\prod_{j=1}^{2} (1-\epsilon_j)^{m_j} \epsilon_j^{N-m_j} \right) \nonumber \\
	&  & \cdot \sum_{\mu} \binom{N}{\mu} \binom{N-\mu}{m_{1}-\mu} \binom{N-m_{1}}{m_{2}-\mu} \mathbb{P}_2(\mathbf{m},\mu;K),\label{eq:Pr_mcast_2users}
\end{eqnarray}
\setlength{\arraycolsep}{5pt}% 
where $\mathbf{m}=(m_1,m_2)$ and the innermost summation is performed over $\mu=\max(0,m_{1} + m_{2}-N),\ldots,\min(m_{1},m_{2})$. Here, $\mu$ denotes the number of common packets received by the two users and $\mathbb{P}_2(\mathbf{m},\mu;K)$ denotes the probability of two correlated random matrices with dimensions $m_1 \times K$ and $m_2 \times K$ and $\mu$ common rows being simultaneously full rank, for $m_1,m_2 \geq K$. This probability is given by
\begin{equation}
	\mathbb{P}_2(\mathbf{m},\mu;K) = \sum_{i} \mathbb{P}^{(i)}(\mu,K) \prod_{j=1}^2 \mathbb{P}(m_{j}-\mu,K-i),\label{eq:ProbFullRankBoth}
\end{equation}
where the summation is performed over the values of $i$ from $\max(0,K-m_{1}+\mu,K-m_{2}+\mu)$ to $\min(\mu,K)$. Term $\mathbb{P}^{(i)}(\mu,K)$ denotes the probability that a random $\mu \times K$ matrix has rank $i$ \cite{Tsimbalo2016a}:
\begin{equation}
	\mathbb{P}^{(i)}(\mu,K) = \frac{1}{q^{(\mu-i)(K-i)}} \prod_{l=0}^{i-1} \frac{(1-q^{l-\mu})(1-q^{l-K})}{1-q^{l-i}}.\label{eq:ProbRanki}
\end{equation} 

The notation introduced in this section is summarised in Table~\ref{tab:Notation}.

\section{Proposed Theoretical Framework\label{sec:Theory}}

We now turn our attention toward the general case of an $L$-user multicast network described at the beginning of the previous section. Our goal is to derive the probability of successful decoding in such a network for two cases - non-systematic and systematic RLNC. We start by formulating a general framework, and then consider each case individually.

The transmission of $N$ coded packets over $L$ lossy links can be modelled as $N$ independent trials. In each trial, the packet can be received by a single user, by a selection of at least two users or by none of the users. The total number of outcomes is equal to $\sum_{i=0}^{L}\binom{L}{i}=2^{L}$. 

{Consider first the packets received by a group of at least two users. Let $\mu$ be a random variable denoting a number of packets received simultaneously by all the users, i.e., \mbox{$\mu=|U_{1}\cap \ldots\cap U_{L}|$}. Furthermore, let \mbox{$\theta_J = \left|\left(\bigcap_{j \in J}U_{j}\right) \cap \left(\bigcap_{j \notin J} \bar{U}_{j}\right) \right|$}, where $J \subset \{1,\ldots,L\}$, \mbox{$1<|J|<L$}, be a random variable denoting the number of transmitted packets received simultaneously by at least two, but less than $L$ users and \emph{not} received by the remaining users. For convenience, let $\theta_J$ obtained for all possible subsets $J$ be assembled in a tuple of $2^L-L-2$ random variables $\boldsymbol{\theta}$.}

{Consider now the packets received by a single user only. However, instead of introducing another set of $L$ random variables, we observe that if the total number of packets received by the $j$-th user, $m_j$, is known, the number of packets received only by this user can be calculated as \mbox{$m_j - \mu  - \sum_{J : j \in J} \theta_J$},
where the summation is performed over all possible subsets $J \subset \{1,\ldots,L\}$, $1<|J|<L$, that include $j$. In other words, the number of packets unique to the $j$-th user is fully determined by $m_j$, $\mu$ and $\boldsymbol{\theta}$.} 

{Finally, the number of packets received by \emph{none} of the users, $\left|\bigcap_{j=1}^{L}\bar{U}_{j}\right|$, can be calculated as follows. If the numbers of packets $m_j$, for $j=1,\ldots,L$, received by each user are added up, the number of packets $\mu$ common to all users will be counted $L$ times. Similarly, the number of packets $\theta_J$ received by a subset of users $J \subset \{1,\ldots,L\}$, $1<|J|<L$, will be counted $|J|$ times. Since each transmitted packet should be counted only once, the number of packets not received by any user can be computed as follows:
\begin{equation}
	\left|\bigcap_{j=1}^{L}\bar{U}_{j}\right| = N - \sum_{j=1}^{L} m_{j} + (L-1)\mu + \sum_{l=2}^{L-1} (l-1) \sum_{J : |J| = l} \theta_J.\label{eq:incl-excl-1} 
\end{equation}
}

\begin{example}
For a multicast network of $L=3$ users,
\begin{itemize}
\item $\mu=|U_{1} \cap U_{2} \cap U_{3}|$,
\item \mbox{$\theta_{\{ 1,2 \}} = |U_{1} \cap U_{2} \cap \bar{U}_{3}|$},
\item \mbox{$\theta_{\{ 1,3 \}} = |U_{1} \cap \bar{U}_{2} \cap U_{3}|$},
\item \mbox{$\theta_{\{ 2,3 \}} = |\bar{U}_{1} \cap U_{2} \cap U_{3}|$}.
\end{itemize}    
The number of packets received uniquely, for instance, by the first user is \mbox{$m_{1} - \mu - \theta_{\{ 1,2 \}} - \theta_{\{ 1,3 \}}$}, 
and the number of packets received by none of the users is \mbox{$N - \sum_{j} m_{j} + 2\mu  + \theta_{\{ 1,2 \}} + \theta_{\{ 1,3 \}} + \theta_{\{ 2,3 \}}$}.
\end{example}

To summarise, the combination of $m_1,\ldots,m_L$, $\mu$ and $\boldsymbol{\theta}$ describes all possible outcomes of the transmission of $N$ coded packets. Let $f(\mathbf{m},\mu,\boldsymbol{\theta};N;\boldsymbol{\epsilon})$ be the joint probability mass function (PMF) of these variables, where $\mathbf{m}=(m_{1},\ldots,m_{L})$.
The PMF can be expressed as follows:
\begin{equation}
	f(\mathbf{m},\mu,\boldsymbol{\theta}; N; \boldsymbol{\epsilon}) = 
		\gamma(\mathbf{m}, \mu, \boldsymbol{\theta};N) \varphi_{L}(\mathbf{m}, N, \boldsymbol{\epsilon}).\label{eq:PMF}
\end{equation}
The first term, $\gamma(\mathbf{m},\mu,\boldsymbol{\theta};N)$, denotes the number of ways to select $\mathbf{m}$, $\mu$ and $\boldsymbol{\theta}$ out of $N$. It can be calculated as a product of binomial coefficients, the number of which is equal to $2^{L}-1$, the total number of elements in $\mathbf{m}$, $\mu$ and $\boldsymbol{\theta}$. 
The second term in \eqref{eq:PMF} denotes the probability of a particular combination of values contained in $\mathbf{m}$, $\mu$, $\boldsymbol{\theta}$ and can be calculated as follows. Consider probability $1-\epsilon_j$, which is associated with coded packets received by the $j$-th user, $j=1,\ldots,L$. The total number of such packets is $m_{j}$. On the other hand, probability $\epsilon_j$ is associated with packets not received by the $j$-th user, the total number of which is $N-m_{j}$. Therefore, $\varphi_{L}(\mathbf{m},N,\boldsymbol{\epsilon})$ can be calculated as follows:
\begin{equation}
	\varphi_{L}(\mathbf{m},N,\boldsymbol{\epsilon}) = \prod_{j=1}^{L} (1-\epsilon_j)^{m_{j}} \epsilon_j^{N-m_{j}}.\label{eq:prob_single_sel}
\end{equation}
We observe that the probability of particular combination of $\mathbf{m}$, $\mu$ and $\boldsymbol{\theta}$ does not depend on $\mu$ or $\boldsymbol{\theta}$.

In general, the probability of successful decoding for a multicast network of $L$ users can be calculated by marginalising the probability of all $L$ coding matrices being full rank over the joint distribution of $\mathbf{m}$, $\mu$ and $\boldsymbol{\theta}$:
\begin{equation}
	P_L(\boldsymbol{\epsilon}) = \sum_{\mathbf{m}, \mu, \boldsymbol{\theta}} 
		f(\mathbf{m},\mu,\boldsymbol{\theta};N;\boldsymbol{\epsilon}) \mathbb{P}_L(\mathbf{m},\mu,\boldsymbol{\theta};K),\label{eq:Pr_mcast_general}
\end{equation}
where
\begin{equation}
	\mathbb{P}_L(\mathbf{m}, \mu, \boldsymbol{\theta};K) = \Pr\left[\bigcap_{j=1}^{L} \mathrm{rank}(\mathbf{C}_{j}) = K\right]\label{eq:prob_n_mat1}
\end{equation}
is the probability that coding matrices $\mathbf{C}_{1},\ldots,\mathbf{C}_{L}$ are simultaneously full rank. 
We note that \eqref{eq:Pr_mcast_general} applies to both non-systematic and systematic codes. One can observe two challenges associated with the direct calculation of \eqref{eq:Pr_mcast_general}. The first challenge is to express the summation over $\mathbf{m}$, $\mu$ and $\boldsymbol{\theta}$ and to calculate the PMF 
$f(\mathbf{m},\mu,\boldsymbol{\theta};N;\boldsymbol{\epsilon})$. Based on the discussion above, the number of nested sums in \eqref{eq:Pr_mcast_general} and the number of binomial coefficients in $f(\mathbf{m},\mu,\boldsymbol{\theta};N;\boldsymbol{\epsilon})$
will grow exponentially with the number of users $L$, thus making the direct approach impractical. 

The second challenge associated with computing \eqref{eq:Pr_mcast_general} is to calculate the probability \eqref{eq:prob_n_mat1} of $L$ correlated matrices being full rank, for a given combination of $\mathbf{m}$, $\mu$ and $\boldsymbol{\theta}$. In Section~\ref{sec:System-model}, \eqref{eq:ProbFullRankBoth} shows how this probability can be exactly calculated for $L=2$ by marginalising it over the distribution of the rank of the submatrix formed by the common rows.
Applying this approach to a larger number of matrices, however, is impractical, since the number of distinct sets of common rows, hence the number of submatrices whose ranks need to be considered, grows exponentially with $L$.

Next, we address the problem of calculating \eqref{eq:Pr_mcast_general} for non-systematic and systematic RLNC.

\subsection{Non-systematic RLNC\label{subsec:Non-sysRLNC}}

We start with the second challenge, the calculation of the probability \eqref{eq:prob_n_mat1} of $L$ correlated coding matrices being full rank. First, we establish the following result:
%
%----------------------------------------------
\begin{lemma}\label{lemma:prob_n_mat_1st_bound}
The probability \eqref{eq:prob_n_mat1} that $L$ correlated random matrices generated over $\mathbb{F}_{q}$ with dimensions $m_{j}\times K$ are full rank, $j=1,\ldots,L$, is lower-bounded as follows:
\setlength{\arraycolsep}{0.1em}
\begin{eqnarray}
	\Pr\left[\bigcap_{j=1}^{L} \mathrm{rank}(\mathbf{C}_{j})=K \right] & \geq & \prod_{j=1}^{L} \Pr \left[\mathrm{rank}(\mathbf{C}_{j}) = K \right]\nonumber\\
	& = &\prod_{j=1}^{L} \mathbb{P}(m_{j}, K),\label{eq:prob_n_mat_lemma1}
\end{eqnarray}
\setlength{\arraycolsep}{5pt}%
where $\mathbb{P}(m_{j},K)$ is given by \eqref{eq:P_single_mat}.
\end{lemma}
%----------------------------------------------
%
\begin{IEEEproof}
See Appendix A.
\end{IEEEproof}

The lower bound \eqref{eq:prob_n_mat_lemma1} is often implicitly used in the literature. For instance, by substituting \eqref{eq:prob_n_mat_lemma1} to \eqref{eq:Pr_mcast_general}, the approximation \eqref{eq:P_M_approx} can be obtained. In contrast with the literature, however, Lemma~\ref{lemma:prob_n_mat_1st_bound} establishes that this approximation is indeed a lower bound.
At the same time, we note that the bound \eqref{eq:prob_n_mat_lemma1} becomes loose if significant correlation between the matrices is present. For instance, consider two matrices with dimensions $m_1\times K$ and $m_2\times K$, such that $m_1\geq m_2\geq K$ and $\mu=m_2$. Clearly, in this case the probability of both matrices having full rank is equal to $\mathbb{P}(m_{2},K)$. However, the same probability as predicted by bound \eqref{eq:prob_n_mat_lemma1} is equal to $\mathbb{P}(m_{1},K)\mathbb{P}(m_{2},K)$, which is smaller than the exact value by $\mathbb{P}(m_{1},K)$ times.

Based on Lemma~\ref{lemma:prob_n_mat_1st_bound}, we now establish a tighter bound for \eqref{eq:prob_n_mat1}:
%
%----------------------------------------------
\begin{lemma}[Improved bound]\label{lemma:prob_n_mat}
The probability \eqref{eq:prob_n_mat1} that $L$ correlated random matrices generated over $\mathbb{F}_{q}$
are full rank is lower-bounded by $\tilde{\mathbb{P}}_L(\mathbf{m},\mu;K)$, which is given by 
\begin{equation}
	\tilde{\mathbb{P}}_L(\mathbf{m}, \mu; K) = \sum_{i} \mathbb{P}^{(i)}(\mu,K) \prod_{j=1}^{L} \mathbb{P}(m_{j}-\mu, K-i),\label{eq:prob_n_mat_main}
\end{equation}
where the summation is performed over the values of $i$ from $\max_{j}(0,K-m_{j}+\mu)$ to $\min(\mu,K)$ 
and $\mathbb{P}^{(i)}(\mu,K)$ is the probability that an $\mu\times K$ matrix has rank $i$, as given by \eqref{eq:ProbRanki}.
\end{lemma}
%----------------------------------------------
%
\begin{IEEEproof}
Using the notation introduced earlier, each of the coding matrices $\mathbf{C}_{1},\ldots,\mathbf{C}_{L}$ has $\mu$ rows common to all of them. By averaging over the distribution of the rank of the matrix formed by these $\mu$ common rows, probability \eqref{eq:prob_n_mat1} for non-systematic RLNC can be expressed as follows:
\setlength{\arraycolsep}{0.1em}
\begin{eqnarray}
	\mathbb{P}_L(\mathbf{m}, \mu, \boldsymbol{\theta}; K) &=& \sum^{\min(\mu, K)}_{i=\max_{j}(0,K - m_{j} + \mu)} \mathbb{P}^{(i)}(\mu, K)\nonumber\\
	& & \cdot \Pr \left[\bigcap_{j=1}^{L} \mathrm{rank}(\mathbf{C}'_{j}) = K-i\right],\label{eq:prob_n_mat0}
\end{eqnarray}
\setlength{\arraycolsep}{5pt}%
where $\mathbf{C}'_{j}$ denotes a matrix formed by the intersection of the $m_{j}-\mu$ rows of $\mathbf{C}_{j}$ not common to all matrices and $K-i$ columns. The starting value of $i$ in the summation in \eqref{eq:prob_n_mat0} is chosen such that for any matrix $\mathbf{C}'_{j}$, there are at least as many rows as columns, i.e., $m_{j}-\mu\geq K-i$. Such starting value excludes unnecessary summation terms. As regards the maximum value of the summation index $i$, it is chosen as the minimum dimension of the matrix formed by $\mu$ common rows. The application of Lemma~\ref{lemma:prob_n_mat_1st_bound} to the second term in the product under the summation in \eqref{eq:prob_n_mat0} results in the lower bound \eqref{eq:prob_n_mat_main}.
\end{IEEEproof}

\begin{remark}
For a two-user multicast network, the bound \eqref{eq:prob_n_mat_main} is exact and reduces to \eqref{eq:ProbFullRankBoth}.
\end{remark}
\vspace{0bp}

We note that by marginalizing over the distribution of the rank of the matrix formed by the rows common to all matrices, the bound \eqref{eq:prob_n_mat_main} is expected to be tighter than that of Lemma \ref{lemma:prob_n_mat_1st_bound}, especially if the number of common rows $\mu$ is large. We illustrate this statement using the following example.

\begin{example}
Consider three $6\times 5$ matrices generated uniformly at random over the binary field $\mathbb{F}_2$, such that all three matrices have $\mu=4$ common rows. Furthermore, assume that each of the three possible pairs of matrices has an additional common row between them. In this case, none of the matrices has rows generated independently from the other matrices.  The probability of all three matrices being full rank estimated by Monte Carlo simulations and obtained by averaging over $10^4$ random realisations is equal to $0.33$. The same probability obtained using the bounds \eqref{eq:prob_n_mat_lemma1} and \eqref{eq:prob_n_mat_main} is equal to $0.20$ and $0.27$, respectively. Clearly, in this example the new bound of Lemma~\ref{lemma:prob_n_mat} provides closer approximation. 
\end{example}

At this point, we have established a more accurate lower bound for the probability \eqref{eq:prob_n_mat1} that all $L$ coding matrices are simultaneously full rank for a given distribution of received packets among the users. We now proceed to the derivation of the probability of successful delivery \eqref{eq:Pr_mcast_general} for the non-systematic case.

%----------------------------------------------
\begin{theorem}\label{theorem:prob_mcast_nsys}
The probability of successful delivery in an $L$-user multicast network characterised by PERs $\boldsymbol{\epsilon}$ and employing an $(N,K,q)$ non-systematic code is lower-bounded as follows: 
\begin{equation}
	P_L(\boldsymbol{\epsilon}) \geq \sum_{\mathbf{m}} \varphi_{L}(\mathbf{m}, N, \boldsymbol{\epsilon})
		\sum_{\mu} \alpha_{L}(\mathbf{m} ,\mu; N) \tilde{\mathbb{P}}_L(\mathbf{m}, \mu; K),\label{eq:pr_mcast_main}
\end{equation}
where
\setlength{\arraycolsep}{0.1em}
\begin{eqnarray}
	\hspace{-2em}
	\alpha_{L}(\mathbf{m}, \mu; N) &=& \binom{N}{\mu} \sum_{l=0}^{\min_{j} (m_{j} - \mu)} (-1)^{l} \binom{N - \mu}{l}\nonumber\\
		& & \hspace{7em} \cdot \prod_{j=1}^{L} \binom{N -\mu - l}{m_{j} - \mu - l}\label{eq:alpha}
\end{eqnarray}
\setlength{\arraycolsep}{5pt}%
and the summation is performed over $m_{j}=K,\ldots,N$ and 
\begin{equation}
	\mu = \max \left(0, \sum_{j=1}^{L}m_{j} - (L-1)N \right), \ldots, \min_{j}m_{j}.
\end{equation}
\end{theorem}
%----------------------------------------------
%
\begin{IEEEproof}
Substituting \eqref{eq:PMF} into \eqref{eq:Pr_mcast_general} and using the fact that $\varphi_{L}$ \eqref{eq:prob_single_sel} does not depend on $\mu$ and $\boldsymbol{\theta}$, the probability in question can be expressed as follows:
\setlength{\arraycolsep}{0.1em}
\begin{eqnarray}
	\hspace{-1em} P_L(\boldsymbol{\epsilon}) & = & \sum_{\mathbf{m}} \varphi_{L}(\mathbf{m}, N, \boldsymbol{\epsilon})\nonumber \\
 		&  & \cdot \sum_{\mu, \boldsymbol{\theta}} \gamma(\mathbf{m}, \mu, \boldsymbol{\theta}; N) \mathbb{P}_L(\mathbf{m}, \mu, \boldsymbol{\theta}; K).\label{eq:Pr_mcast_general-1}
\end{eqnarray}
\setlength{\arraycolsep}{5pt}%
We now employ Lemma~\ref{lemma:prob_n_mat} and bound \eqref{eq:Pr_mcast_general-1} from below:
\setlength{\arraycolsep}{0.1em}
\begin{eqnarray}
	P_L(\boldsymbol{\epsilon}) & \geq & \sum_{\mathbf{m}} \varphi_{L}(\mathbf{m}, N, \boldsymbol{\epsilon}) \sum_{\mu} \tilde{\mathbb{P}}_L(\mathbf{m}, \mu; K)\nonumber \\
 		&  & \cdot \sum_{\boldsymbol{\theta}} \gamma(\mathbf{m}, \mu, \boldsymbol{\theta}; N).\label{eq:Pr_mcast_general-1-1}
\end{eqnarray}
\setlength{\arraycolsep}{5pt}%
To prove \eqref{eq:pr_mcast_main}, we now show that the innermost sum in \eqref{eq:Pr_mcast_general-1-1} is equal to $\alpha_{L}(\mathbf{m},\mu;N)$ given by \eqref{eq:alpha}. To this end, we rewrite this sum as follows:
\begin{equation}
	\sum_{\boldsymbol{\theta}} \gamma(\mathbf{m}, \mu, \boldsymbol{\theta}; N) = \binom{N}{\mu} \beta,
\end{equation}
where
\begin{equation}
	\beta = \sum_{\boldsymbol{\theta}} \gamma(\mathbf{m} - \mu, 0, \boldsymbol{\theta}; N - \mu).\label{eq:beta}
\end{equation}
In other words, by selecting $\mu$ out of $N$ packets common to all users, $\beta$ is the total number of possible selections of
$\sum_{j=1}^{L}(m_{j}-\mu)$ packets out of $N-\mu$ such that none of the packets is received by all users at the same time. The value
of $\beta$ can be calculated by the inclusion-exclusion principle in its complementary form \cite{VanLint}. To this end, let $S$ denote a set of all possible selections of $\sum_{j=1}^{L}(m_{j}-\mu)$ packets out of $N-\mu$. The number of elements in this set is 
\begin{equation}
	|S| = \prod_{j=1}^{L} \binom{N - \mu}{m_{j} - \mu}.
\end{equation}
Consider now subsets $S_{k}$ of $S$, $k=1,\ldots,N-\mu$, containing selections corresponding to the $k$-th transmitted packet received
by all the $L$ users. Let $\bar{S}_{k}$ denote the complement of $S_{k}$ in $S$. It can be observed that $\beta$ can be thought of as the
cardinality of a set constructed as the intersection of all $\bar{S}_{k}$, $k=1,\ldots,N-\mu$. Using the inclusion-exclusion principle, 
\begin{eqnarray}
	\beta & = & |\bigcap_{k = 1}^{N - \mu} \bar{S}_{k}|\nonumber \\
 		  & = & |S| - \sum_{k = 1}^{N - \mu} |S_{k}| + \sum_{1 \leq k_{1} < k_{2} \leq N-\mu} |S_{k_{1}} \cap S_{k_{2}}| - \ldots \nonumber \\
 		  &   & -(-1)^{z} \sum_{1 \leq k_{1} < \ldots < k_{z} \leq N-\mu} |S_{k_{1}} \cap \ldots \cap S_{z}|,\label{eq:incl-excl}
\end{eqnarray}
where $z=\min_{j}(m_{j}-\mu)$ is the minimum possible number of packets received by all the $L$ users. The first summation in \eqref{eq:incl-excl} corresponds to $N-\mu$ possible selections of a single packet received by all the users, with $\prod_{j=1}^{L}\binom{N-\mu-1}{m_{j}-\mu-1}$ selections for other $N-\mu-1$ available packets. Similarly, the second summation in (\ref{eq:incl-excl}) corresponds to $\binom{N-\mu}{2}$ selections of two commonly received packets and $\prod_{j=1}^{L}\binom{N-\mu-2}{m_{j}-\mu-2}$ selections of other packets. Expression \eqref{eq:incl-excl} can therefore be written in a compact form as follows:
\begin{equation}
\beta=\sum_{l=0}^{z}(-1)^{l}\binom{N-\mu}{l}\prod_{j=1}^{L}\binom{N-\mu-l}{m_{j}-\mu-l},\label{eq:beta1}
\end{equation}
thus making the inner sum in \eqref{eq:Pr_mcast_general-1-1} equal to $\alpha_{L}(\mathbf{m},\mu;N)$ \eqref{eq:alpha}. 

The values of $(m_{1},\ldots,m_{L})$ over which the outer-most summation in (\ref{eq:pr_mcast_main}) is performed are chosen so
that each user should receive at least $K$ packets. As regards the number of packets $\mu$ received by all users, its maximum value
cannot exceed the smallest $m_{j}$, for $j=1,\ldots,L$. The starting value of $\mu$ can be found assuming that all other $N-\mu$ transmitted packets have been simultaneously received by $L-1$ users. As a result, $\mu \geq \sum_{j}m_{j}-(L-1)N.$ If $(L-1)N>\sum_{j}m_{j}$, the starting value of $\mu$ should be $0$.
\end{IEEEproof}

\begin{remark}\label{rem:rem1_mcast_nonsys}
We note that bound \eqref{eq:pr_mcast_main} is obtained by applying Lemma~\ref{lemma:prob_n_mat}, meaning that only packets received simultaneously by all users are considered to take into account the correlation effect. {As a result, the bound is expected to be especially tight if the number of such packets $\mu$ is likely to be large, which is typical in scenarios where a non-negligible fraction of users experiences PERs that are relatively small. We observe that this is the case of multicast networks where users are spread across the coverage area of the source node (namely, a base station serving a cell). As argued in~\cite{7248795}, 3GPP's LTE-A systems~\cite{LTE} are likely to ensure reduced user PERs across the majority of the cell area.} Finally, for large values of PER, $\mu$ is likely to be small and the bound converges to the traditional approximation \eqref{eq:P_M_approx}.  
\end{remark}

\begin{remark}\label{rem:rem2_mcast_nonsys}
The derived bound is exact for $L=2$ users and matches \eqref{eq:Pr_mcast_2users}.
Indeed, the product of binomial coefficients in \eqref{eq:Pr_mcast_2users} can be shown to be equal to $\alpha_{2}(\mathbf{m},\mu;N)$ defined by \eqref{eq:alpha} as follows. Without loss of generality, let $m_{1}\leq m_{2}$. The last binomial coefficient in \eqref{eq:Pr_mcast_2users} is equivalent to the number of $N-m_{2}$ selections out of $N-\mu$ packets, such that each selection includes $m_{1}-\mu$ packets. We can again employ the inclusion-exclusion principle and denote $S$ as a set of all possible selections, with the number of elements in this set equal to $\binom{N-\mu}{N-m_{2}}$. Let $S_{k}$ denote a subset of $S$ containing selections in which the $k$-th packet belonging to the group of $m_{1}-\mu$ packets is not included, $k=1,\ldots,m_1-\mu$. The last binomial coefficient in \eqref{eq:Pr_mcast_2users} can be expressed as follows:
\setlength{\arraycolsep}{0.14em} 
\begin{eqnarray}
	\hspace{-1em} \binom{N - m_{1}} {m_{2} - \mu} & = & \left| \bigcap_{k=1}^{m_{1} - \mu} \bar{S}_{k} \right| \nonumber \\
 		& = & \sum_{l=0}^{m_{1} - \mu} (-1)^{l} \binom{m_{1} - \mu}{k} \binom{N - \mu - l}{N - m_{2}}.\label{eq:incl-excl-2}
\end{eqnarray}
\setlength{\arraycolsep}{5pt}%
Multiplying the right-hand side of \eqref{eq:incl-excl-2} with the first two binomial coefficients of \eqref{eq:Pr_mcast_2users}
yields:
\setlength{\arraycolsep}{0.14em} 
\begin{eqnarray*}
	\binom{N}{\mu} \binom{N - \mu}{m_{1} - \mu} \sum_{l=0}^{m_{1} - \mu} (-1)^{l} \binom{m_{1} - \mu}{l} \binom{N - \mu - l}{N - m_{2}}\\
	= \binom{N}{\mu} \sum_{k=0}^{m_{1} - \mu} (-1)^{l} \binom{N - \mu - l}{m_{1} - \mu - l} \binom{N - \mu - l}{m_{1} - \mu - l}\\
	= \alpha_{2}(\mathbf{m}, \mu; N).
\end{eqnarray*}
\setlength{\arraycolsep}{5pt}%
\end{remark}

\subsection{Systematic RLNC}\label{sec:sysRLNC}

As pointed out in Remark~\ref{rem:rem1_mcast_nonsys}, the bound \eqref{eq:pr_mcast_main} derived for non-systematic RLNC was obtained by considering packets received simultaneously by all users, thus partially taking into account the correlation between their coding matrices. In the case of systematic RLNC, the correlation arises only from commonly received \emph{non-systematic} packets, since the systematic packets correspond to deterministic vectors of coding coefficients. For small values of PER, each user is likely to receive all $K$ systematic packets regardless of the number of received non-systematic packets. Even for large values of PER, the correlation effect is smaller than in the case of non-systematic RLNC, since the number of transmitted non-systematic packets for systematic RLNC is smaller than the total number of transmissions. Therefore, it is expected that for systematic RLNC, considering a multicast network as a set of independent unicast connections will result in an approximation close enough for any field size and PER. We now state this result formally and prove that such approximation is a lower bound, as in the case of non-systematic RLNC.

%----------------------------------------------
\begin{theorem}\label{th:theorem_sysRLNC}
The probability of successful delivery $P_L^*(\boldsymbol{\epsilon})$ of the $L$-user multicast network characterised by PERs $\boldsymbol{\epsilon}$ and employing an $(N,K,q)$ systematic code is lower-bounded as follows:
\begin{equation}
	P_L^*(\boldsymbol{\epsilon}) \geq \prod_{j=1}^{L} P^*(\epsilon_j),\label{eq:theorem_sys}
\end{equation}
where $P^*(\epsilon_j)$ is the probability of successful delivery of a point-to-point link with a PER $\epsilon_j$ given by \eqref{eq:P_ptp_sys}.
\end{theorem}
%----------------------------------------------

\begin{IEEEproof}
Consider the general formulation for the probability of successful delivery of the multicast network  \eqref{eq:Pr_mcast_general}, but with marginalisation over the distribution of $\mathbf{m}$ only:
\begin{equation}
	P_L^*(\boldsymbol{\epsilon}) = \sum_{\mathbf{m}} \varphi_{L}(\mathbf{m}, N, \boldsymbol{\epsilon}) 
		\mathbb{P}_L^*(\mathbf{m}, K) \prod_{j=1}^{L} \binom{N}{m_j},\label{eq:theorem_sys_proof_1}
\end{equation}
where $\mathbb{P}_L^*(\mathbf{m}, K)$ denotes the probability that all $L$ coding matrices are simultaneously full rank. This probability can be marginalised over the probability distribution of the number of systematic packets $h_j$ received by the $j$-th user as follows:
\setlength{\arraycolsep}{0.14em}
\begin{eqnarray}
	\mathbb{P}_L^*(\mathbf{m}, K) &=& \sum_{h_1} \ldots \sum_{h_L} \left[ \prod_{j=1}^{L} \frac{\binom{K}{h_j} \binom{N-K}{m_j-h_j}}{\binom{N}{m_j}} \right] \nonumber\\ 
		& & \cdot \Pr \left[ \bigcap_{j=1}^{L} \mathrm{rank}(\mathbf{C}'_{j}) = K - h_j \right],\label{eq:theorem_sys_proof_2}
\end{eqnarray}\setlength{\arraycolsep}{5pt}%
where matrix $\mathbf{C}'_{j}$ is composed of the intersection of $m_j-h_j$ non-systematic rows and $K-h_j$ columns of the $j$-th coding matrix $\mathbf{C}_{j}$. Equation \eqref{eq:theorem_sys_proof_2} can be lower-bounded by applying Lemma~\ref{lemma:prob_n_mat_1st_bound} to its probability term and employing the distributive law  as follows:
\begin{equation}
	\mathbb{P}_L^*(\mathbf{m}, K) \geq \prod_{j=1}^{L} \sum_{h_j} \frac{\binom{K}{h_j} \binom{N-K}{m_j-h_j}}{\binom{N}{m_j}} \mathbb{P}(m_j-h_j,K-h_j).\label{eq:theorem_sys_proof_3} 
\end{equation}
The bound \eqref{eq:theorem_sys} can now be obtained by substituting \eqref{eq:theorem_sys_proof_3} into \eqref{eq:theorem_sys_proof_1}.
\end{IEEEproof}

Bound \eqref{eq:theorem_sys} will be investigated in Section~\ref{sec:Res_sys}, where we will show that it is sufficiently tight even for binary codes and small values of PER.

\subsection{Computational Complexity Consideration}\label{sec:Complexity}

Comparing bounds \eqref{eq:pr_mcast_main} and \eqref{eq:theorem_sys} for non-systematic and systematic codes, it can be observed that the former is significantly more complex than the latter, especially if the number of users $L$ is large.
In this section, we show how the calculation of bound \eqref{eq:pr_mcast_main} in the non-systematic case can be optimised.

Consider the outermost summation in \eqref{eq:pr_mcast_main}, which is performed over $L$ variables contained in tuple $\mathbf{m}$, with each variable taking values from $K$ to $N$. As a result, the number of terms in the summation is equal to $(N-K+1)^L$, which makes \eqref{eq:pr_mcast_main} computationally prohibitive if the number of users $L$ is large. At the same time, it can be observed that $\alpha_{L}(\cdot)$ and $\mathbb{P}_L(\cdot)$ in \eqref{eq:pr_mcast_main}, the most computationally intensive terms, do not depend on the order of elements within $\mathbf{m}$. Thus, the number of times these terms are calculated can be significantly reduced as follows. 

Let us rewrite relation \eqref{eq:pr_mcast_main} as follows:
\setlength{\arraycolsep}{0.14em}
\begin{eqnarray}
	\hspace{-1em} P_L(\boldsymbol{\epsilon}) &\geq & \sum_{\mathbf{m}'} 
			\left( \sum_{\pi(\mathbf{m}')} \varphi_{L}(\mathbf{m}, N, \boldsymbol{\epsilon}) \right) \nonumber\\
		& & \cdot \sum_{\mu} \alpha_{L}(\mathbf{m}', \mu; N) \tilde{\mathbb{P}}_L(\mathbf{m}', \mu; K),\label{eq:pr_mcast_main_opt1}
\end{eqnarray}
\setlength{\arraycolsep}{5pt}%
where the outer summation is now performed over all possible combinations $\mathbf{m}'$ of $L$ values from $K$ to $N$ \emph{with no reference to order}, and $\pi(\mathbf{m}')$ denotes a permutation of $\mathbf{m}'$. The problem of calculating the number of possible combinations of $L$ values from $K$ to $N$ can be recast as that of finding the number of ways to place $L$ balls into $N-K+1$ urns. Indeed, a particular combination $\mathbf{m}'$ is equivalent to one way of assigning each of the $N-K+1$ values (urns) a non-negative number of occurrences (balls) $L_i$, $i=1,\ldots,N-K+1$, so that
\begin{equation}
	\sum_{i=1}^{N-K+1} L_i = L.\label{eq:occup_problem}
\end{equation} 
The number of solutions to \eqref{eq:occup_problem}, and hence the number of terms in the outer summation in \eqref{eq:pr_mcast_main_opt1}, can be found using the stars and bars principle \cite{Feller} and is equal to
\begin{equation}
	\binom{N-K+1+L-1}{L} = \binom{N-K+L}{L}.\label{eq:occup_problem2}
\end{equation}
This number is much smaller than $(N-K+1)^L$, the number of terms in the outer summation in the original expression \eqref{eq:pr_mcast_main}, as illustrated in the following example.

\begin{example}
Let $L=4$ and $N-K=10$. Based on \eqref{eq:occup_problem2}, the number of terms in the outer summation in \eqref{eq:pr_mcast_main_opt1} is equal to $\binom{14}{4}=1001$, which is $14.6$ times smaller than $11^4=14641$, the number of terms in the outer summation in the original expression \eqref{eq:pr_mcast_main}. As a result, the number of times $\alpha_{L}(\cdot)$ and $\mathbb{P}_L(\cdot)$ are calculated is significantly reduced compared with the original expression. Clearly, the reduction factor will increase with $L$.
\end{example}

Further complexity reduction can be achieved in the case of a homogeneous network, in which each user has the same PER $\epsilon=\epsilon_j$ for $j=1,\ldots,L$. In this case, \eqref{eq:pr_mcast_main_opt1} simplifies to
\setlength{\arraycolsep}{0.14em}
\begin{eqnarray}
	\hspace{-1em} P_L(\epsilon) &\geq & \sum_{\mathbf{m}'} \varphi_{L}'(\mathbf{m}', N, \epsilon) \sigma(\mathbf{m}') \nonumber\\
		& & \cdot \sum_{\mu} \alpha_{L}(\mathbf{m}', \mu; N) \tilde{\mathbb{P}}_L(\mathbf{m}', \mu; K),\label{eq:pr_mcast_main_opt2}
\end{eqnarray}
\setlength{\arraycolsep}{5pt}%
where 
\begin{equation}
	\varphi_{L}'(\mathbf{m}',N,\epsilon) = (1-\epsilon)^{\sum \mathbf{m}'} \epsilon^{LN-\sum \mathbf{m}'}
\end{equation}
and $\sigma(\mathbf{m}') = \sum_{\pi(\mathbf{m}')}$ is the number of permutations of a particular combination $\mathbf{m}'$ of $L$ values from $K$ to $N$. This number depends on a particular solution to \eqref{eq:occup_problem} and can be calculated as follows:
\begin{equation}
	\sigma(\mathbf{m}') = \frac{L!}{\prod_{i=1}^{N-K+1} L_i!}.
\end{equation}

To further speed up calculation, probabilities $\mathbb{P}(\cdot)$ and $\mathbb{P}^{(i)}(\cdot)$, which are used repetitively in the calculation of $\tilde{\mathbb{P}}_L(\cdot)$, can be pre-computed offline for a given $(N,K,q)$ code and stored in a look-up table.

\section{Numerical Results\label{sec:NumResults}}

In this section, we investigate the performance of a multicast network under non-systematic and systematic RLNC via simulation and compare the results with the derived theoretical bounds. Simulation results in terms of probability of successful delivery were obtained using the Kodo C++ network coding library \cite{Pedersen2011} and the Monte Carlo method, with each point being the result of an average over $10^{5}$ iterations. The results are compared for various combinations of network and code parameters, as summarised in Table~\ref{tab:Parameters}. The accuracy of the bounds is evaluated in terms of Mean Square Error (MSE) for a given combination of $L$, $\boldsymbol{\epsilon}$, $K$ and $q$  over the range of numbers of transmissions $N$. Unless otherwise stated, a homogeneous scenario is assumed, in which each user experiences the same PER $\epsilon$.

\begin{table}
\begin{centering}
\caption{Network and code parameters used to evaluate the accuracy of the proposed bounds.}
\begin{tabular}{|c|p{3.05cm}|}
\hline 
Parameter & Values\tabularnewline
\hline 
\hline 
Number of users $L$ & $\left\lbrace 2,6,10\right\rbrace$\tabularnewline
\hline 
PER $\epsilon$ & $\left\lbrace 0.01,0.1\right\rbrace$\tabularnewline
\hline 
Number of source packets $K$ & $\left\lbrace 5,10,15,20\right\rbrace$\tabularnewline
\hline 
Number of transmissions $N$ & $\left\lbrace K,K+1,\ldots,K+10\right\rbrace$\tabularnewline
\hline 
Finite field size $q$ & $\left\lbrace 2,2^8\right\rbrace$\tabularnewline
\hline
\end{tabular}
\par\end{centering}
\label{tab:Parameters}

\end{table}

\subsection{Non-systematic RLNC}\label{sec:Res_nsys}

\begin{figure}[t]
\subfloat[$\epsilon=0.01$]{\label{fig:res1a_nsys}
	\includegraphics[width=1\columnwidth]{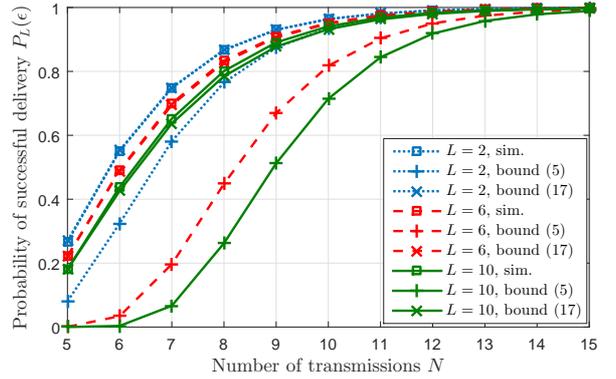}
}\\
\subfloat[$\epsilon=0.1$]{\label{fig:res1b_nsys}
	\includegraphics[width=1\columnwidth]{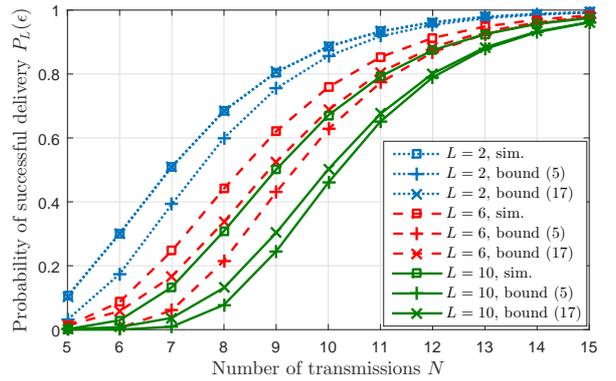}
}\\
\caption{Probability of successful delivery for a binary non-systematic code as a function of $N$ for $K=5$ and $L\in \left\lbrace 2,6,10\right\rbrace$.}
\label{fig:res1_nsys}
\end{figure}

We start with a multicast network operating under a binary, non-systematic code corresponding to $q=2$. Specifically, we compare the proposed bound \eqref{eq:pr_mcast_main} against bound \eqref{eq:P_M_approx} used in the literature for large field sizes. In addition, we benchmark both bounds against simulated results.

Fig.~\ref{fig:res1_nsys} shows the probability of successful delivery $P_L(\epsilon)$ to $L\in \left\lbrace 2,6,10\right\rbrace$ users as a function of the number of coded transmissions $N$. The number of source packets is fixed to $K=5$ and two PER values common to all users are considered: $\epsilon=0.01$ and $0.1$. The latter value of $\epsilon$ is commonly used in practice as the maximum acceptable PER. For instance, in 3GPP's LTE-A systems, the link adaptation mechanism typically switches the modulation and coding scheme once the transport block error rate reaches $0.1$ \cite{LTE,6883864}. Therefore, this value of PER can be thought of as a worst-case scenario. It can be observed that the proposed bound as per \eqref{eq:pr_mcast_main} provides better approximation than bound \eqref{eq:P_M_approx}. Specifically, bound \eqref{eq:pr_mcast_main} is particularly accurate for $\epsilon=0.01$, where it matches the simulated results for $L=2$ ($\mathrm{MSE}=2 \cdot 10^{-6}$) and closely follows them when $L\in \left\lbrace 6,10\right\rbrace$ ($\mathrm{MSE}=8 \cdot 10^{-6}$ and $9 \cdot 10^{-5}$, respectively). The tightness of the proposed bound in this scenario is explained by a large number of packets likely to be received simultaneously by all users. Indeed, when $\epsilon$ is small, the probability that a single transmitted packet is received by all users, which is equal to $(1-\epsilon)^L$, is large. This leads to a high correlation between the users coding matrices. This is in contrast with the traditional bound \eqref{eq:P_M_approx}, which does not take into account commonly received packets. As a result, bound \eqref{eq:P_M_approx} is particularly loose when $\epsilon$ is small, with the absolute gap from the simulated results being up to $0.58$ when $L=10$ and $N=7$, in contrast with $0.016$ for the new bound. At the same time, as $\epsilon$ or $L$ increase, the probability that a packet will be received by all users decreases, so the accuracy of the proposed bound decreases too, as can be observed for $\epsilon=0.1$ and $L=10$ ($\mathrm{MSE}=0.01$). We reiterate, however, that bound \eqref{eq:pr_mcast_main} is exact for $L=2$ and any $\epsilon$, as per Remark~\ref{rem:rem2_mcast_nonsys}.

\begin{figure}[t]
\subfloat[$\epsilon=0.01$]{\label{fig:res2a_nsys}
	\includegraphics[width=1\columnwidth]{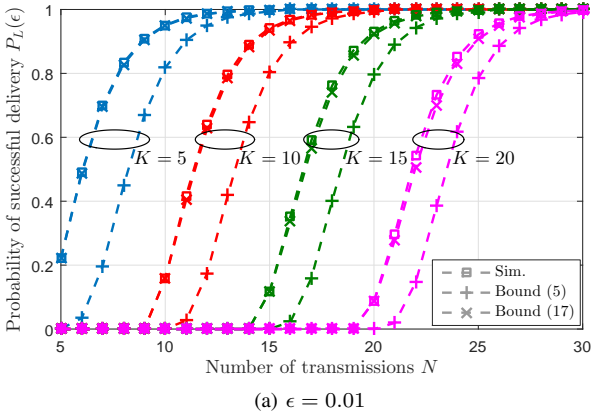}
}\\
\subfloat[$\epsilon=0.1$]{\label{fig:res2b_nsys}
	\includegraphics[width=1\columnwidth]{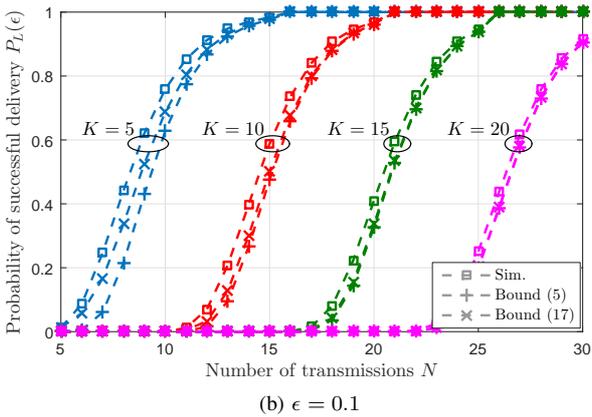}
}\\
\caption{Probability of successful delivery for a binary non-systematic code as a function of $N$ for $L=6$ and $K\in \left\lbrace 5,10,15,20\right\rbrace$.}
\label{fig:res2_nsys}
\end{figure}

Fig.~\ref{fig:res2_nsys} illustrates the results for the same scenario as in Fig.~\ref{fig:res1_nsys}, but this time for different numbers of source packets $K\in \left\lbrace 5,10,15,20\right\rbrace$ and a fixed number of users $L=6$. In line with the previous results, the proposed bound \eqref{eq:pr_mcast_main} closely follows the simulated performance at $\epsilon=0.01$, exhibiting the MSE of $8 \cdot 10^{-6}$, $6 \cdot 10^{-5}$, $1 \cdot 10^{-4}$ and $3 \cdot 10^{-4}$ for $K\in \left\lbrace 5,10,15,20\right\rbrace$, respectively. The new bound is significantly more accurate than bound \eqref{eq:P_M_approx} at $\epsilon=0.01$, with the latter having the MSE of up to $7 \cdot 10^{-2}$. It can be observed that for this PER, the accuracy of the proposed bound somewhat decreases as $K$ grows. The reason is that longer source messages require more coded transmissions, which leads to a higher number of packet erasures for a given PER. As a result, the correlation between the users coding matrices reduces for larger $K$, which means a smaller number of packets received simultaneously by all users. For the same reason, it can be observed from Fig.~\ref{fig:res2_nsys} that the gap between the simulated results and bound \eqref{eq:P_M_approx} becomes smaller as $K$ grows. For example, for $\epsilon=0.01$, the maximum gap between bound \eqref{eq:P_M_approx} and the simulated results reduces from $0.5$ for $K=5$ ($N=7$) to $0.4$ for $K=20$ ($N=22$). Fig.~\ref{fig:res2b_nsys} demonstrates that both bounds are close to the simulated results for the worst-case scenario in terms of PER, especially when $K=20$ -- $\mathrm{MSE}=9 \cdot 10^{-4}$ and $8 \cdot 10^{-4}$ for bounds \eqref{eq:P_M_approx} and \eqref{eq:pr_mcast_main}, respectively. It should be noted, however, that due to the high decoding complexity of non-systematic RLNC \cite{Lucani2010}, $K$ is likely to be small in practice, and in this regime the proposed bound \eqref{eq:pr_mcast_main} is noticeably more accurate ($\mathrm{MSE}=3 \cdot 10^{-3}$ when $K=5$) than \eqref{eq:P_M_approx} ($\mathrm{MSE}=1.4 \cdot 10^{-2}$ for the same value of $K$) even if $\epsilon$ is high.

\begin{figure}[t]
\includegraphics[width=1\columnwidth]{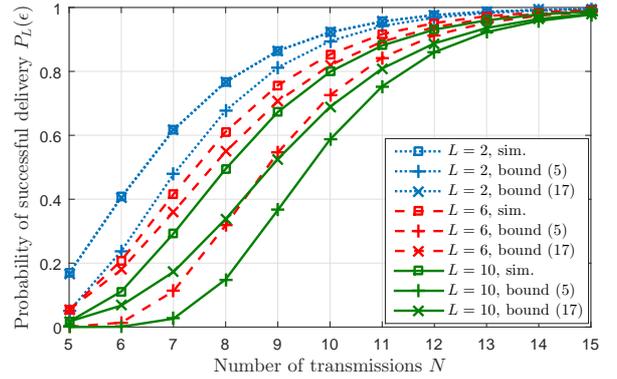}
\caption{Probability of successful delivery for a binary non-systematic code as a function of the number of transmissions $N$ for the case when each user has a unique PER from $0.01$ to $0.1$, for $L\in \left\lbrace 2,6,10\right\rbrace$ and $K=5$.}
\label{fig:res1_het_nsys}
\end{figure}

The results so far were collected for a homogeneous network, in which each user has the same PER $\epsilon$. It is also relevant to verify the performance of the bounds for a more general case of a heterogeneous network, in which users have distinct PERs. To this end, we allocate $L$ unique PER values from $0.01$ to $0.1$ with a constant step equal to $0.99/(L-1)$, which is equivalent to a set of users placed on the symmetry axis of a cellular cell sector \cite{Tassi2015}. Fig.~\ref{fig:res1_het_nsys} compares the bounds and simulated results for such network, for the same values of $L$ and $K$ as in Fig.~\ref{fig:res1_nsys}. It is clear that the proposed bound \eqref{eq:pr_mcast_main} is significantly more accurate than \eqref{eq:P_M_approx} even for $L=10$, with the maximum absolute gap between the two bounds being $0.17$, $0.24$ and $0.19$ for $L=2$ $(N=6)$, $L=6$ $(N=7)$ and $L=10$ $(N=8)$, respectively.
{\begin{remark}
Comparing with Fig.~\ref{fig:res1_nsys}, it can be observed that the gap between the proposed bound and simulation results is larger than when each user has $\epsilon=0.01$, but smaller than when each user has $\epsilon=0.1$. This is an expected result, since the users have varying PERs between those two values. Still, we observe that the correlation effect given by the number $\mu$ of packets received simultaneously by all the users is relevant and accounting for this makes our bound~\eqref{eq:pr_mcast_main} tighter than~\eqref{eq:P_M_approx}.
\end{remark}}

\begin{figure}[t]
\includegraphics[width=1\columnwidth]{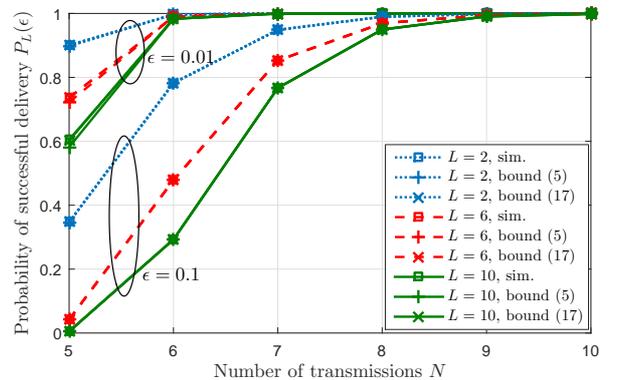}
\caption{Probability of successful delivery for a non-binary ($q=2^8$), non-systematic code as a function of the number of transmissions $N$ for $L\in \left\lbrace 2,6,10\right\rbrace$, $\epsilon\in \left\lbrace 0.01,0.1 \right\rbrace$ and $K=5$.}
\label{fig:res1_q256_nsys}
\end{figure}

Finally, Fig.~\ref{fig:res1_q256_nsys} compares the bounds and simulated results for a non-binary, non-systematic code. A relatively large field size, $q=2^8$, is selected, in which the traditional bound \eqref{eq:P_M_approx} is expected to be accurate. It can be observed that while bound \eqref{eq:P_M_approx} closely follows the simulated performance when $L=2$ ($\mathrm{MSE}=10^{-6}$), it somewhat diverges from the simulated results at small $N$ for a larger number of users and $\epsilon=0.01$: for $N=5$, the absolute gap is $0.01$ and $0.02$ for $L=6$ and $10$, respectively. It is expected that the deviation of bound \eqref{eq:P_M_approx} will grow further when $L$ is increased beyond $10$. In other words, bound \eqref{eq:P_M_approx}, traditionally used in the literature, has a noticeable approximation error, that grows with $L$, even for a large field size. By contrast, the proposed bound \eqref{eq:pr_mcast_main} is clearly more accurate when $\epsilon=0.01$, with the largest deviation from the simulated results (corresponding to $L=10$ and $N=5$) being $2 \cdot 10^{-3}$, $10$ times lower than that of bound \eqref{eq:P_M_approx}. As regards $\epsilon=0.1$, both bounds accurately describe the performance, with the worst-case MSE (corresponding to $L=10$) equal to $8 \cdot 10^{-4}$ and $7 \cdot 10^{-4}$ for bound \eqref{eq:P_M_approx} and \eqref{eq:pr_mcast_main}, respectively.

To summarise, the proposed bound \eqref{eq:pr_mcast_main} for non-systematic RLNC is tighter than the existing bound \eqref{eq:P_M_approx} traditionally used in the literature. The difference between the bounds is especially profound in realistic channel conditions, when users have small values of PER.    

\subsection{Systematic RLNC}\label{sec:Res_sys}
\begin{figure}[t]
\includegraphics[width=1\columnwidth]{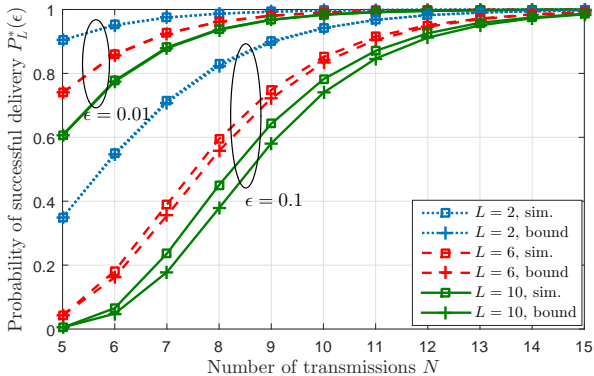}
\caption{Probability of successful delivery for a binary systematic code as a function of the number of transmissions $N$ for $L\in \left\lbrace 2,6,10\right\rbrace$, $\epsilon\in \left\lbrace 0.01,0.1 \right\rbrace$ and $K=5$.}
\label{fig:res1_sys}
\end{figure}

We now turn our attention to a multicast network operating under systematic RLNC, the performance bound for which was proposed in Theorem~\ref{th:theorem_sysRLNC}. In this scenario, the first $K$ transmissions are the original source packets, followed by coded, non-systematic packets. Fig.~\ref{fig:res1_sys} compares bound \eqref{eq:theorem_sys} with simulated performance for various numbers of users $L$ and PER $\epsilon$. The number of source packets $K$ is fixed to $5$. It can be observed that the bound is accurate when $\epsilon=0.01$, with the MSE of $4 \cdot 10^{-7}$, $7 \cdot 10^{-7}$ and $2 \cdot 10^{-6}$ for $L=2$, $6$ and $10$, respectively. This is in contrast with the same bound applied to non-systematic RLNC \eqref{eq:P_M_approx}, which exhibited significant inaccuracy at small PER. Such behaviour of bound \eqref{eq:theorem_sys} at small $\epsilon$ was predicted in Section~\ref{sec:sysRLNC} and explained by the high probability that each user receives all $K$ source packets. Even when $\epsilon=0.1$ and $L=10$, the vertical gap between the bound and simulated results is relatively small: up to $7 \cdot 10^{-2}$ for $N=8$, compared to $0.58$ in the non-systematic case. This phenomenon was again predicted in Section~\ref{sec:sysRLNC} and is due to the fact that the correlation between the users coding matrices arises from the non-systematic packets only, the number of which is smaller than the total number of transmissions.
Due to the tightness of bound \eqref{eq:theorem_sys} for both considered PER values, the results for a heterogeneous scenario, in which every user has a distinct PER, are omitted.

\begin{figure}[t]
\includegraphics[width=1\columnwidth]{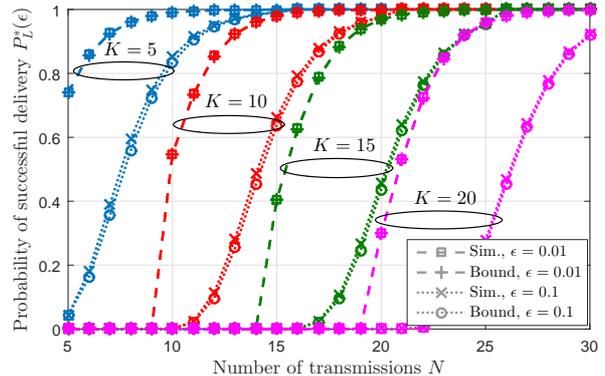}
\caption{Probability of successful delivery for a binary systematic code as a function of the number of transmissions $N$ for $L=6$, $\epsilon\in \left\lbrace 0.01,0.1 \right\rbrace$ and $K\in \left\lbrace 5,10,15,20 \right\rbrace$.}
\label{fig:res2_sys}
\end{figure}

By analogy to Fig.~\ref{fig:res2_nsys} for non-systematic RLNC, Fig.~\ref{fig:res2_sys} illustrates the performance results for the systematic case, for fixed $L=6$ and variable $K$, for two values of PER $\epsilon$. The bound provides close approximation for all $K$ at $\epsilon=0.01$: the MSE ranges from $7 \cdot 10^{-7}$ (for $K=5$) to $5 \cdot 10^{-6}$ (for $K=20$). At the same time, a small deviation can be observed when $\epsilon=0.1$, which decreases as $K$ becomes larger -- the MSE drops from $4 \cdot 10^{-4}$ (for $K=5$) to $9 \cdot 10^{-5}$ for ($K=20$). The deviation can be explained by a larger influence of the correlation between the users coding matrices when $\epsilon$ is high, which is not taken into account by the proposed bound. The deviation reduction as $K$ increases is due to a decreasing amount of correlation, as in the case of non-systematic RLNC. 

\begin{figure}[t]
\includegraphics[width=1\columnwidth]{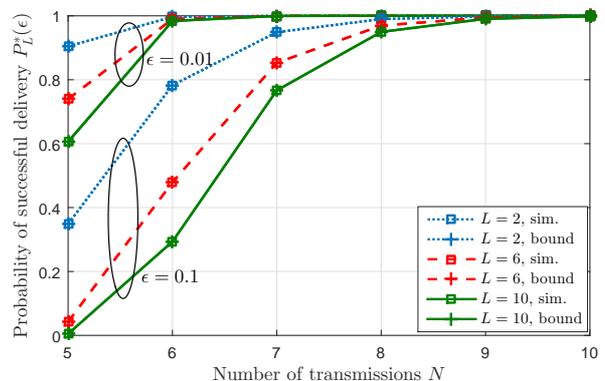}
\caption{Probability of successful delivery for a non-binary $q=2^8$, systematic code as a function of the number of transmissions $N$ for $L\in \left\lbrace 2,6,10\right\rbrace$, $\epsilon\in \left\lbrace 0.01,0.1 \right\rbrace$ and $K=5$.}
\label{fig:res1_q256_sys}
\end{figure}

Finally, Fig.~\ref{fig:res1_q256_sys} demonstrates the performance results for a systematic non-binary code, corresponding to $q=2^8$. The accuracy of the bound can be observed for all considered values of $L$ and $\epsilon$, with the worst-case MSE (corresponding to $L=10$) equal to $4 \cdot 10^{-7}$ and $2 \cdot 10^{-7}$, for $\epsilon=0.01$ and $0.1$, respectively. We note that the results in Fig.~\ref{fig:res1_q256_sys} are similar to those shown in Fig.~\ref{fig:res1_q256_nsys} for a non-systematic code. Indeed, when the field size is large, the probability of successful delivery in both cases can be closely approximated by the probability of each user collecting at least $K$ packets, which does not depend on the nature of the code.

All in all, it can be concluded that for systematic RLNC, bound \eqref{eq:theorem_sys} is sufficiently tight for most considered combinations of $K$, $L$, $\epsilon$ and $q$, with a small deviation occurring when $K=5$, $L \geq 6$ and $\epsilon=0.1$. 

\section{Conclusions and Future Work\label{sec:Conclusions}}

In this paper, we have addressed the issue of calculating the probability of successful delivery in a multicast network operating under RLNC. In contrast with the previous studies focused on specific network or code parameters, we considered the most general scenario of arbitrary number of users and finite field size, as well as realistic channel conditions. In addition to the traditional, non-systematic form of RLNC, we have also considered the systematic version.

For non-systematic RLNC, we proposed a novel lower bound for the probability of successful delivery, which takes into account a limited finite field size and potential correlation between the users. This is in contrast with the bound traditionally used in the literature, which assumes an infinite or sufficiently large field size and independence between subsets of packets received by each user. {For systematic RLNC, however, we argued that the correlation effect between the users is negligible, and the traditionally used bound is sufficiently tight.

The accuracy of the considered bounds was thoroughly investigated via Monte Carlo simulations for various combinations of network and code parameters.} In the non-systematic scenario, it was demonstrated that the proposed bound is significantly more accurate than the traditional bound used in the literature. The accuracy of the new bound was shown to be especially high at low PER, exhibiting an MSE of $9 \cdot 10^{-5}$ for a ten-user network. In contrast, the absolute deviation of the state-of-the art bound is as large as $0.58$ for the same network. Even for a large finite field, it was shown that the traditional bound deviates from the simulated results, while the proposed bound is up to $10$ times more accurate. {In particular, this holds true in scenarios where the multicast users experience heterogenous PERs.} {In the systematic case, the considered bound was shown to be sufficiently tight for most configurations, with a small deviation occurring at high PER and small message size.} By examining the accuracy of the bounds, we provided a unique insight into the selection of code parameters for various network configurations.

The considered bounds for the probability of successful delivery could be used to obtain other performance metrics, such as the average decoding delay or energy efficiency. In addition, the derived results can be utilised in the analysis of other network topologies, such as relay networks. To improve the utility of the bounds, a further reduction in their complexity, especially in the non-systematic case, can be investigated.

\section*{Appendix A}

\section*{Proof of Lemma~\ref{lemma:prob_n_mat_1st_bound}}

Let $A_{j}$, $j=1,\ldots,L$, denote an event corresponding to matrix $\mathbf{C}_{j}$ being full rank. Using this notation, \eqref{eq:prob_n_mat_lemma1} can be rewritten as
\begin{equation}
	\Pr \left[ \bigcap_{j=1}^{L} A_{j} \right] \geq \prod_{j=1}^{L} \Pr \left[ A_{j} \right].\label{eq:prob_n_mat}
\end{equation}

We first show that \eqref{eq:prob_n_mat} is valid for $L=2$. Consider the two matrices in question, $\mathbf{C}_{1}$ and $\mathbf{C}_{2}$. Using the notation of Section \ref{sec:Theory}, let $\mu$ denote the number of common rows in these matrices. Let also $\mathbf{X}$ denote a matrix formed by those rows. The joint probability of both matrices being full rank can be expressed as follows:
\begin{equation}
	\Pr[A_1 \cap A_2] = \sum_{i} \Pr[\mathrm{rank}(\mathbf{X}) = i, A_1 \cap A_2],\label{eq:prob_n_mat3}
\end{equation}
where the maximum value of the summation index $i$ is limited by $\min(\mu,K)$.
Let $Z$ be a random variable representing the rank of $\mathbf{X}$, so that both matrices are full rank. Therefore, \eqref{eq:prob_n_mat3} can be thought of as the Cumulative Distribution Function (CDF) of $\mu$, $F(\mu)$:
\begin{equation}
	\Pr[A_1 \cap A_2] = \sum_{i}\Pr[Z = i] = \Pr[Z \leq \mu].
\end{equation}
Since the CDF is a non-decreasing function, it follows that its minimum value corresponds to $\mu = 0$, for which the matrices are independent from each other. Hence,
\begin{equation}
	\Pr[A_1 \cap A_2] \geq F(0) = \Pr[A_1] \Pr[A_2],\label{eq:prob_n_mat4}
\end{equation}
which proves \eqref{eq:prob_n_mat} for $L=2$.

For $L>2$, the left-hand side of \eqref{eq:prob_n_mat} can be expressed based on the chain rule as follows:
\setlength{\arraycolsep}{0.14em}
\begin{eqnarray}
	\Pr \left[ \bigcap_{j=1}^{L} A_{j} \right] & = & \Pr \left[ A_{L} \biggl | \bigcap_{j=1}^{L-1} A_{j} \right]
			\cdot \Pr \left[ A_{L-1} \biggl | \bigcap_{j=1}^{L-2} A_{j} \right] \cdot \ldots \nonumber\\
 		& & \cdot \Pr \left[ A_{2} | A_{1} \right] \cdot \Pr[A_{1}] \nonumber\\
 		&=& \prod_{l=1}^{L} \Pr \left[A_{l} \biggl | \bigcap_{j=1}^{l-1} A_{j} \right].\label{eq:chain_rule}
\end{eqnarray}
\setlength{\arraycolsep}{5pt}%
From \eqref{eq:prob_n_mat4}, it follows that
\begin{equation}
	\Pr[A_2|A_1] = \frac{\Pr[A_1 \cap A_2]}{\Pr[A_1]} \geq \Pr[A_2]. 
\end{equation}

Consider now the term in the product \eqref{eq:chain_rule} corresponding to $l=3$, $\Pr[A_{3}|A_{1} \cap A_{2}]$. It can be expressed as follows:
\begin{equation}
	\Pr[A_{3}|A_{1} \cap A_{2}] = \frac{\Pr[A_3 \cap A_2 | A_1]}{\Pr[A_2 | A_1]}.\label{eq:pr_l3}
\end{equation}
From \eqref{eq:prob_n_mat4}, it follows that
\begin{equation}
	\Pr[A_{3} \cap A_{2} | A_{1}] \geq \Pr[A_3 | A_1]\Pr[A_2 | A_1].
\end{equation}
Substituting this into \eqref{eq:pr_l3} leads to
\begin{equation}
\Pr[A_{3}|A_{1} \cap A_{2}] \geq \Pr[A_3 | A_1] \geq \Pr[A_3].
\end{equation} 

Using the same logic, it is straightforward to show that if 
$\Pr[A_{l-1} | \bigcap_{j=1}^{l-2}A_{j}] \geq \Pr[A_{l-1}]$ holds, so does 
$\Pr[A_{l} | \bigcap_{j=1}^{l-1}A_{j}] \geq \Pr[A_{l-1}]$. Indeed,
\begin{eqnarray}
	\Pr \left[ A_{l} | \bigcap_{j=1}^{l-1} A_{j} \right] = \frac{\Pr \left[ A_{l} \cap A_{l-1} | \bigcap_{j=1}^{l-2} A_{j} \right]}
															   {\Pr \left[ A_{l-1} | \bigcap_{j=1}^{l-2} A_{j} \right]} \nonumber\\
		\geq \Pr \left[ A_l | \bigcap_{j=1}^{l-2} A_{j} \right] \geq \Pr[A_l].
\end{eqnarray}
As a result, every term in the product \eqref{eq:chain_rule} is lower-bounded by the corresponding marginal probability, which leads to \eqref{eq:prob_n_mat} and proves the lemma.$\hfill\blacksquare$

\bibliographystyle{ieeetr}
\bibliography{TCOM_Multicast_RLNC}

\end{document}